\documentclass[]{spie}  %>>> use for US letter paper
\usepackage{array}
\usepackage{wrapfig}
\usepackage{multirow}
\usepackage{tabularx}
\usepackage{cite}
\usepackage{multicol}
\usepackage{balance}
\usepackage[colorlinks=true, allcolors=blue]{hyperref}
\usepackage[pdftex]{graphicx}
\usepackage{epstopdf}
\usepackage{subfigure}

\usepackage{pgfplots,multicol}
\usepackage{amsmath,stmaryrd,pict2e,picture}
\usepackage{amsmath,amssymb}
\usepackage{mathtools}
\usepackage{pifont}% http://ctan.org/pkg/pifont

\newlength{\rowidth}% row operation width
\AtBeginDocument{\setlength{\rowidth}{3em}}
\usepackage{mathrsfs}
\usepackage{algorithm, algorithmic}

\def\footnoterule{\relax%
	\kern-5pt
	\hbox to \columnwidth{\hfill\vrule width 0.75\columnwidth height 0.4pt\hfill}
	\kern4.6pt}
\makeatother
\title{Radar Human Motion Classification Using Multi-Antenna System}

\author{Patrick~A~Schooley}
\author{Syed~A.~Hamza}
\affil{School of Engineering, Widener University, Chester, PA 19013, USA}

\authorinfo{E-mails: 	 \{paschooley, shamza\}@widener.edu}

\begin{document}

\maketitle
\begin{abstract}
This paper considers human activity classification for an indoor radar system.  Human motions generate nonstationary radar returns which represent Doppler and micro-Doppler signals. The time-frequency (TF) analysis of micro-Doppler signals can discern subtle variations on the motion by precisely revealing velocity components of various moving body parts. We consider radar for activity monitoring using TF-based machine learning approach exploiting both temporal and spatial degrees of freedom. The proposed approach captures different human motion representations more vividly in joint-variable data domains achieved through beamforming at the receiver. The radar data is collected using real time measurements at 77 GHz using four receive antennas, and subsequently micro-Doppler signatures are analyzed  through machine learning algorithm for  classifications of  human walking motions. We present the performance of the proposed multi antenna approach in separating and classifying two closely walking persons moving in opposite directions.
\end{abstract}
%\begin{IEEEkeywords}
%MIMO radar, sparse array transceiver, adaptive beamforming, group sparsity, %successive convex approximation
%\end{IEEEkeywords}

%We consider two separate scenarios namely, both the transmit/receive can potentially share one or more common sensor locations or  the transmit/receive perspective sensor locations are disjoint. 
%\IEEEpeerreviewmaketitle
\section{Introduction}

Human motion recognition (HMR) finds important applications in a large variety of scenarios ranging from gesture recognition for smart homes, detecting events of interest for automatic  surveillance, behavioral analysis, Gait abnormality recognitions, health monitoring in  care facilities and  rehabilitation services   to enable independent living for elderly \cite{6126543, Amin2017RadarFI, 8610109, WANG2018118, 8613848, 9069251}.

Contactless  sensing of  human motions  has gained traction because of the obvious  benefits of being non obtrusive. It does not require any user intervention and as such  the users are not required to wear specific devices to be tracked  via smart phone applications  \cite{7426551, 10.1117/12.2527660, 8894733}. Radar systems are at the forefront of remote sensing technologies as they provide robust non  contact monitoring that is   not affected by  lighting conditions. Additionally, active RF (radio frequency) sensing provides 4D imaging capabilities by explicitly measuring the scatterer velocity in addition to range and 2D angular localization.  This is unlike other remote sensing sensors of human motions such as visual-based systems that require additional pre-processing and filtering operations to accurately discern small movements \cite{8010417, 8746862}. Also radar images are privacy preserving  as high resolution imaging radar   renders silhouette type portrait revealing little identifiable information as opposed to camera-based systems.

We investigate a human activity monitoring system to concurrently monitor movements of multiple persons.  This could facilitate to separately record the activities of  multiple persons in detail. For example, in health care facility,  the task of care givers can be eased by attending to the needs of  several care receivers at the same time. Radar is a proven technology for  target detection, localization and tracking. Imaging radars are getting attention recently because of their added capability of   classifying different targets. Radar returns classification could be performed after localizing the target in range, azimuth and/or Doppler.  

In this paper, we consider the radar human motion classification by attempting to localize the motions to a given azimuth directions. The proposed beamforming approach can reduce the system cost and alleviate the need of using multiple radars as proposed in \cite{8010417, 8835796, 8835618,  article123}. Azimuth filtering is achieved by  applying  beamforming to the receiver array. Specifically, we consider the task of classifying two persons walking closely in opposite directions at different azimuth angles. We aim to correctly  pair the direction of motion to  the corresponding azimuth angle. This is achieved by jointly processing  the two spectrograms obtained  in the  directions of both motions. In this case, the received data is filtered using beamforming with two separate sets of beamformer weights. Subsequently, classification is performed by jointly processing the time frequency signature of both azimuth directions.   The proposed scheme works adequately  when the two motions are not completely separable in azimuth. This could either be due to the close proximity of the two motions in the azimuth or high sidelobes of the beamformer.   

\begin{figure*}[!t]
	\centering
	\includegraphics[height=1.9in, width=6.5in]{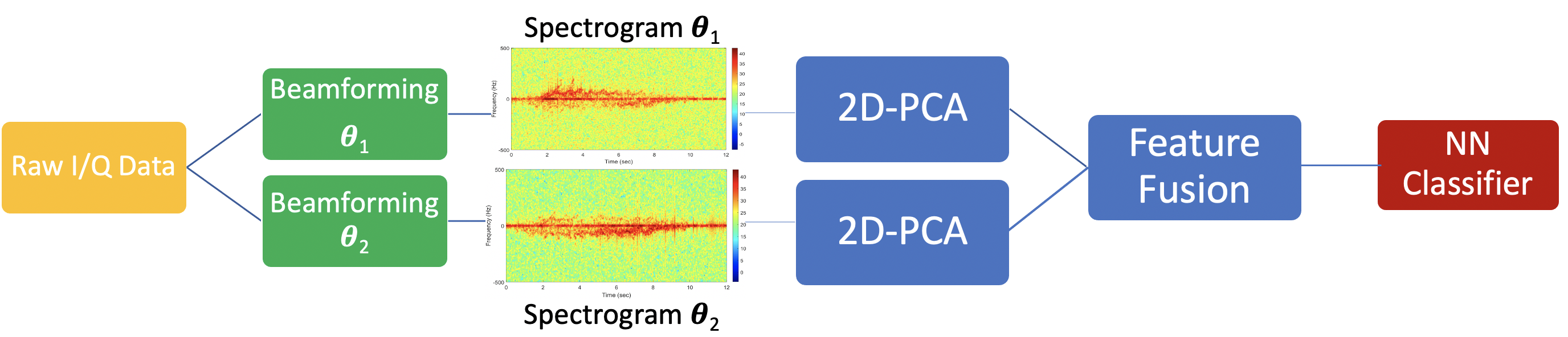}
	\caption{Processing sequence}
	\label{Fig.1}
\end{figure*} 
The rest of the paper is organized as follows:  In the next section, we  describe the  data set and time frequency representation domain. The 2D PCA and beamformer image fusion  for multi-person monitoring is discussed in Section \ref{Feature extraction and classification}. Experimental results are shown in Section \ref{Experimental Results}, while the conclusion is given in Section \ref{Conclusion}.
%Section \ref{Optimum sparse array design} elaborates on the  sparse array design by  semidefinite relaxation to jointly find the optimum sparse transmit/receive array geometry. In the subsequent section,  simulations are presented to demonstrate the offerings of the proposed sparse array transmit design. The paper ends with concluding remarks.
% We demonstrate the offerings of the proposed sparse array design approach by comparing its performance with those    sparse arrays  developed by  existing design methods. \par
\section{Radar return signal analysis}
\label{Problem Formulation}
The complex valued raw  data matrix $\mathbf{s}(n,m) \in C^{N*M}$  of the frequency-modulated continuous
wave (FMCW) radar  is obtained through spatially processing the radar returns    by an  $M$ element uniformly spaced antenna array. The data is collected over  $N$ temporal sampling instances. The receiver array vector $\mathbf{s}(m) \in C^{M}$ at  time instant $n$ corresponds to the $n_{th}$ row of $\mathbf{s}(n,m)$ and is given by,
\begin {equation} \label{a}
\mathbf{s}(m)=    \sum_{l=1}^{L} \alpha _l \mathbf{a}( \theta_l)  + \mathbf{v}(m),
\end {equation}
where, $\mathbf{a} ({\theta_l})$  $\in \mathbb{C}^{M}$ is  the  steering vector   corresponding to the azimuth direction $\theta_l$ of the scatterer, and is defined  as follows,  
\vspace{+2mm}
\begin {equation}  \label{b}
\mathbf{a} ({\theta_l})=[1 \,  \,  \, e^{j (2 \pi / \lambda) d cos(\theta_l)  } \,  . \,   . \,  . \, e^{j (2 \pi / \lambda) d (M-1) cos(\theta_l)  }]^T.
\end {equation}
Here, $d$ is the inter-element spacing  and $\alpha_l$ $\in \mathbb{C}$  is the complex amplitude of the radar return. The additive Gaussian noise $\mathbf{v}(m)$ $\in \mathbb{C}^M$ has   variance   $\sigma_v^2$.
The elements of the received data vector $\mathbf{s}(m)$   are   combined linearly by the $M$-sensor  beamformer that strives to spatially filter the reflections from all other directions except  the signal in the direction of beamformer look angle $\theta_k$. The spatially filtered signal vector $\mathbf{x}({\theta_k})$ $\in \mathbb{C}^N$ after beamforming is given by, 
\begin {equation}  \label{c}
\mathbf{x}({\theta_k}) = \mathbf{s}(n,m) \mathbf{w}^H({\theta_k}),
\end {equation}
where $\mathbf{w}({\theta_k})=\mathbf{a}^H ({\theta_k})$ are the complex beamformer weights pointing towards $\theta_k$.

The spatially filtered signal vector $\mathbf{x}({\theta_k})$  is reshaped into a two-dimensional matrix, $\mathbf{x}_{\theta_k}(p, q)$. This is achieved  by segmenting the $N$ dimensional vector $\mathbf{x}({\theta_k})$, such that, the $P$ samples collected within a pulse repetition interval (PRI) are stacked into a $P$ dimensional column. There are   $Q$ such columns  within $\mathbf{x}_\theta(p, q)$ where  $Q=N/P$  is the number of PRIs processed within the observation time $N$.  The range-map, $\mathbf{r}_{\theta_k}(p,q)$  is obtained by applying the column-wise Discrete Fourier Transform (DFT) operation which is given by,
\begin {equation}  \label{c}
\mathbf{r}_{\theta_k}(l,q) =  \sum_{p=0}^{P-1} \mathbf{x}_{\theta_k}(p, q)e^{-j (2 \pi l p/ N)}
\end {equation}
We observe the data in the TF domain after localizing the motion in azimuth and range bins of interest. The spectrogram is used as the TF signal representation, showing the variation of the signal power as a function of  time $n$ and frequency $k$. The spectrogram of a periodic version of
a discrete signal $\mathbf{v}_{\theta_k}(n)$, is given by \cite{30749, article4573,  9101078}, 
\begin {equation}  \label{c}
\mathbf{d}_{\theta_k}(n,k) = | \sum_{m=0}^{H-1} \mathbf{h}(m)\mathbf{v}_{\theta_k}(n-m)e^{-j (2 \pi k m/ H)}|^2,
\end {equation}
where $\mathbf{v}_{\theta_k}=\sum_{l=r_l}^{r_u}\mathbf{r}_{\theta_k}(l,q)$ is obtained by collapsing the range dimension beginning  from lower range bin $r_l$ to highest range bin $r_h$. Tapering window $\mathbf{h}$ of length $H$  is applied to reduce the sidelobes. The spectrograms reveal the the different velocities, accelerations and higher order moments which cannot be easily modeled or assumed to follow specific nonstationary structures \cite{10.1117/12.669003}. We observe  the  motion of two humans walking closely in opposite directions at different azimuth angles. We aim to correctly  pair the direction of motion to  the corresponding azimuth angle. This is achieved by jointly processing  two spectrograms,  $\mathbf{v}_{\theta_1}(n,k)$ and $\mathbf{v}_{\theta_2}(n,k)$ which are respectively localized at azimuth angles $\theta_1$ and $\theta_2$.    It is clear that the concurrent motions of multiple objects are hard to  be  distinguished in azimuth by  only using a single antenna. 
%For example, falling and sitting
%can be confused in the TF domain, while they can be readily
%distinguished in the range map.two % window overlapping. The spectrogram is resized with 128 samples for Doppler scaling and 32 samples (1 sec) in slow-time. The same resizing process is applied for range-map images

\section{Feature extraction and classification}
\label{Feature extraction and classification}
We adopt Two-Dimensional Principal Component Analysis (2-D PCA) for dimensionality reduction to draw the most pertinent features from spectrograms \cite{8835840, 1360091}. The  features obtained from  individual spectrograms are jointly   classified with the Nearest Neighbor (NN) classifier \cite{9114613}.

The 2-D PCA is performed on the covariance matrix $\mathbf{R}_{\theta_k}$ which is computed as follows,
\begin {equation}  \label{3c}
\mathbf{R}_{\theta_k} =  \sum_{i=0}^{T-1} {\mathbf{X}^{i}}^H_{\theta_k}\mathbf{X}^i_{\theta_k},
\end {equation}
where, $\mathbf{X}^i_{\theta_k}$  is the normalized spectrogram for the $i_{th}$ example and $T$ are the total training examples.  The eigendecomposition of $R_{\theta_k}$ is performed and the individual train images are projected onto the subspace spanned by the $K$ dominant eigenvectors of $R_{\theta_k}$ corresponding to the $K$ largest eigenvalues.  It is noted that  two spectrogram images are generated  per example, by using two different sets of beamformer weights.  After separately performing the  2-D PCA,  the projected spectrograms are vectorized and concatenated  before training the NN classifier, as shown in Fig. \ref{Fig.1}.

The overall data collection, preprocessing and 
classification  can be described by the following steps. 
\subsection{Data Collection and Preprocessing}

\begin{itemize}
 \item  PRI is set to 1 ms, and each data example is observed over the time period of 12 s, resulting in $Q=12000$ slow time samples.
 \item  ADC sampling rate is 512 ksps, rendering 512 fast time samples per PRI, resultantly the length of data vector is $N=6144000$. 
 \item  The  received data $\mathbf{s}(n,m) \in C^{N*M}$, is collected through $M=4$ element receive array, with an inter-element spacing of $\lambda/2$ ($\lambda$ is the wavelength corresponding to the operating frequency), therefore the dimensionality of received raw data matrix is $6144000\times4$. 
 \item  Beamforming is performed on the raw data matrix, resulting in a spatially filtered $\mathbf{x}({\theta_k})$ vector of dimensions $6144000\times1$.  Two such vectors are generated in the directions of each motion ${\theta_1}$ and ${\theta_2}$.
 \item  Each vector $\mathbf{x}({\theta_k})$
is reshaped into a $512\times12000$ matrix. After applying columnwise DFT,  and identifying the range bins of interest, the corresponding rows are summed together, resulting in  $\mathbf{v}_{\theta_k}=\sum_{l=r_l}^{r_u}\mathbf{r}_{\theta_k}(l,q)$, which is of dimensions $12000\times1$. 
 \item  Two spectrogram $\mathbf{d}_{\theta_1}$ and $\mathbf{d}_{\theta_2}$, each of dimensions $384\times128$ is obtained, where the window length is 128. 
 \end{itemize}

\subsection{Training}
\begin{itemize}

\item Normalize both spectrograms of all training examples  $\mathbf{v}_{\theta_k}$ by subtracting the average. 
\item Perform the eigendecomposition of the covariance
matrices estimated according to (\ref{3c}) and select $K$ dominant eigenvectors. Note that  covariance matrices are separately evaluated for both angles.
\item Project the training spectrograms onto the subspace spanned by selected eigenvectors, resulting in two projected spectrograms per example. Each has the dimensionality $128\times K$. 
\item  After vectorizing each spectrogram, both are concatenated into a $256K\times1$  vector and trained through NN classifier.
 \end{itemize}
\subsection{Testing}
\begin{itemize}
\item Normalize the testing image and project onto the subspace obtained in
the training process. 
\item The testing image  is then passed onto trained NN classifier  for prediction. 
 \end{itemize}

\begin{table}[h]
\centering
\caption{Confusion matrix, Number of principal eigenvectors, $K$=1}
\label{table:1}
\begin{tabular}{ | m{10em} | m{2cm}| m{2cm} | } 
\hline
Classified/Actual Class& Class-1 & Class-2 \\ 
\hline
Class-1 & 97.6\% & 0\% \\ 
\hline
Class-2 & 2.4\%  & 100\%  \\ 
\hline  
%\title{Principle Eigen Vector 1}
\end{tabular} 
%\end{center}
\end{table} 
\hfill \break
\vspace{3mm}
\begin{table}[h]
\centering
\caption{Confusion matrix, Number of principal eigenvectors, $K$=2}
\label{table:2}
\begin{tabular}{ | m{10em} | m{2cm}| m{2cm} | } 
\hline
Classified/Actual Class& Class-1 & Class-2 \\ 
\hline
Class-1 & 100\%  & 0\%  \\ 
\hline
Class-2 & 0\%  & 100\%  \\ 
\hline
\end{tabular}
\end{table}

 \section{Experimental Results} \label{Experimental Results}
In this section, we demonstrate the effectiveness of beamformer image fusion  for monitoring multiple persons in the field of view. We  consider two classes, Class-1 and Class-2,  of motions. For Class-1 motions,  one person  moves radially towards the radar at an azimuth angle of $\theta_1$ while at the same time another person moves radially away from the radar at an azimuth angle of $\theta_2$.  For  Class-2 motion, the two persons perform the same walking motion but in an opposite direction, i.e., this time the person at azimuth angle $\theta_1$   moves radially away from the radar, whereas  the  other person concurrently moves radially towards the radar at an azimuth angle of $\theta_2$.

\begin{figure}[!t]
	\centering
	\includegraphics[height=2.3in, width=3.5in]{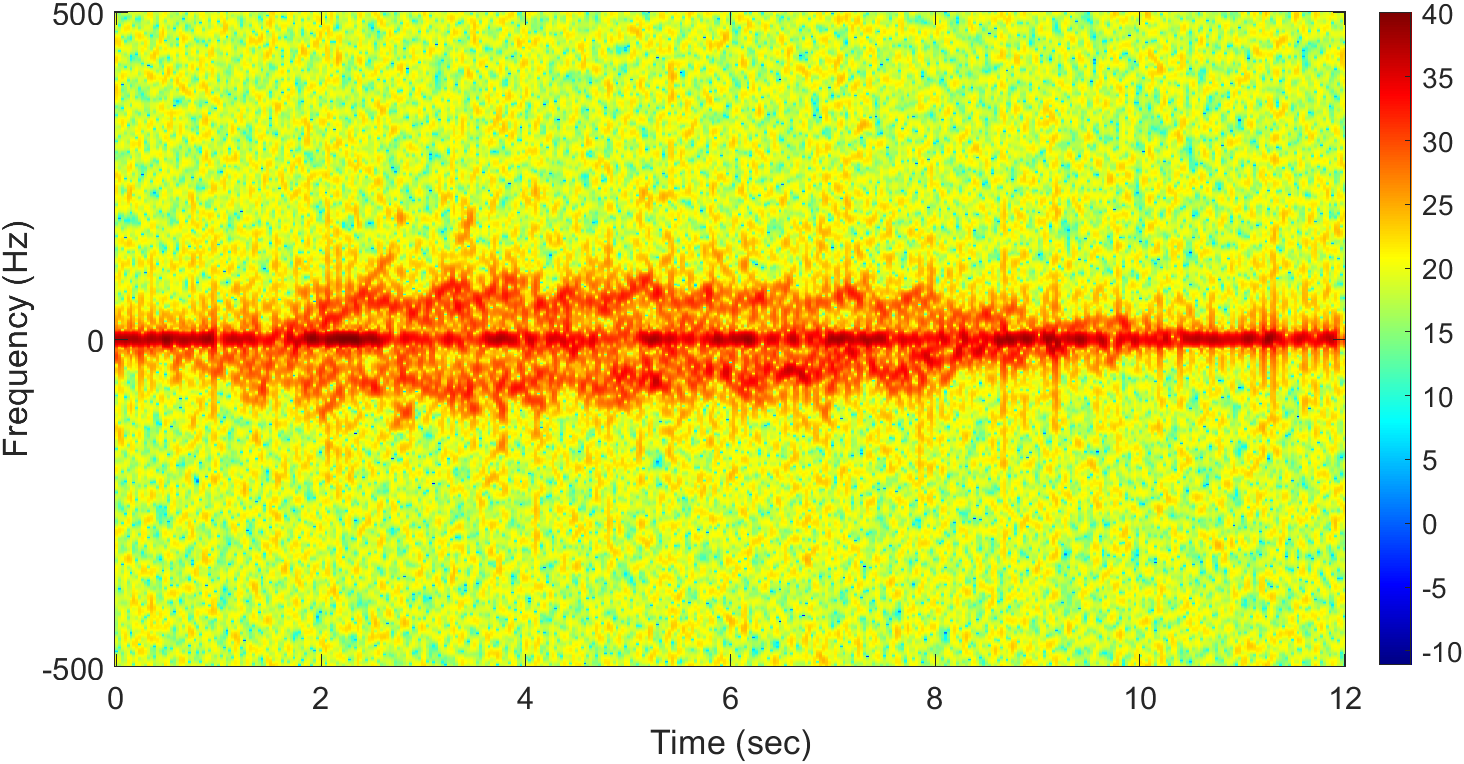}
	\caption{Time Frequency signature of Class-1 motions processed through single antenna (without beamforming)}
	\label{Fig.2}
\end{figure}

\begin{figure}[!t]
	\centering
	\includegraphics[height=2.3in, width=3.5in]{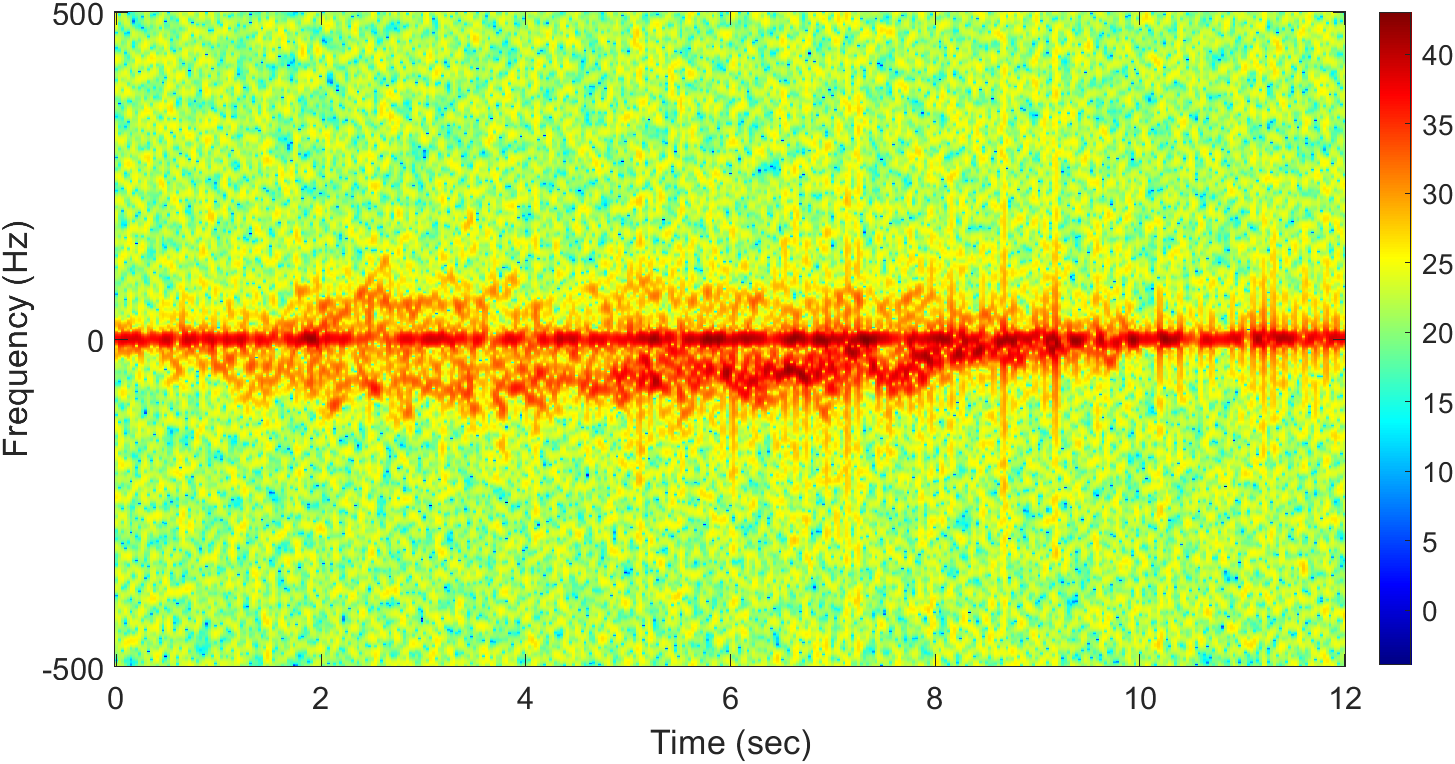}
	\caption{Time Frequency signature of Class-1 motions processed through beamformer pointing towards $\theta_1$.}
	\label{Fig.3}
\end{figure}

\begin{figure}[!t]
	\centering
	\includegraphics[height=2.3in, width=3.5in]{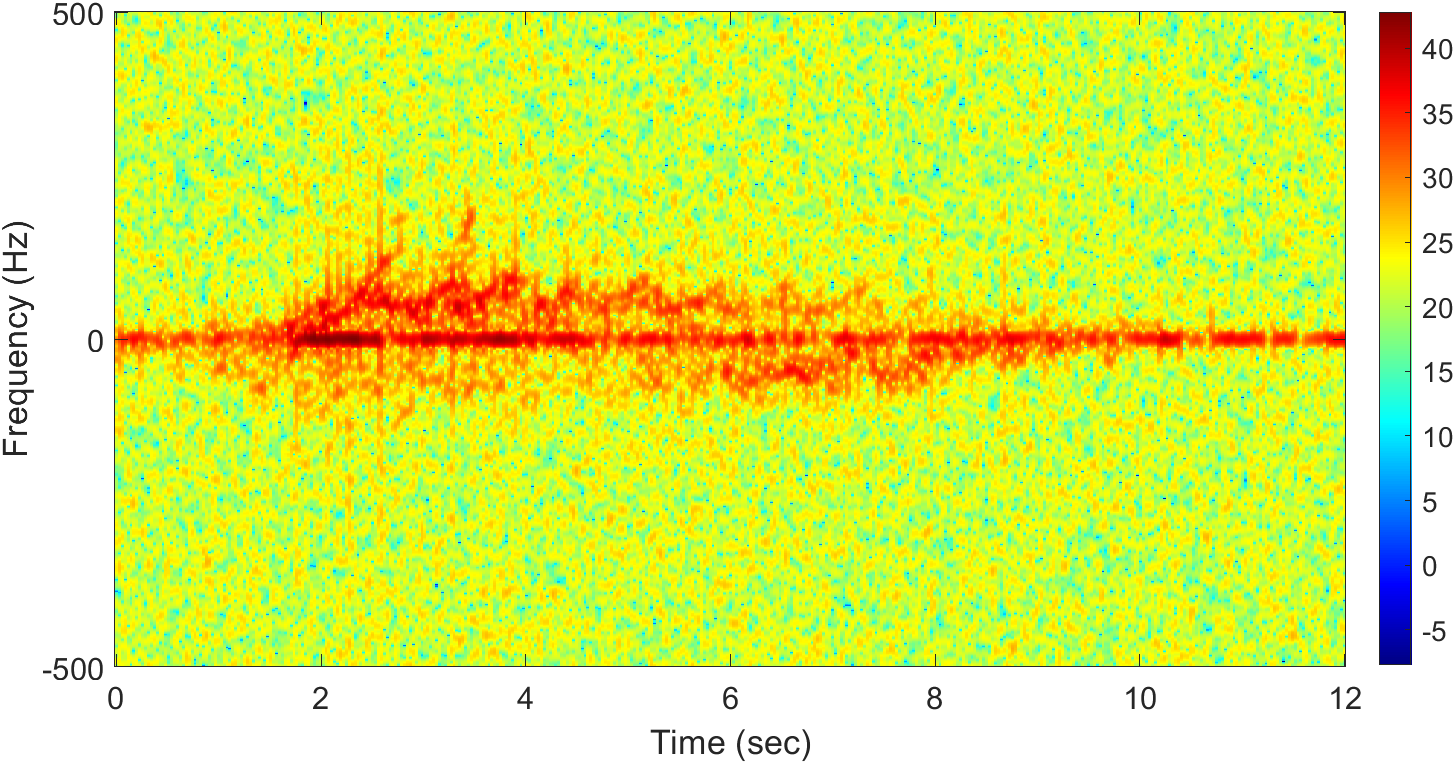}
	\caption{Time Frequency signature of Class-1 motions processed through   beamformer pointing towards $\theta_2$.}
	\label{Fig.4}
\end{figure}

\begin{figure}[!t]
	\centering
	\includegraphics[height=2.3in, width=3.5in]{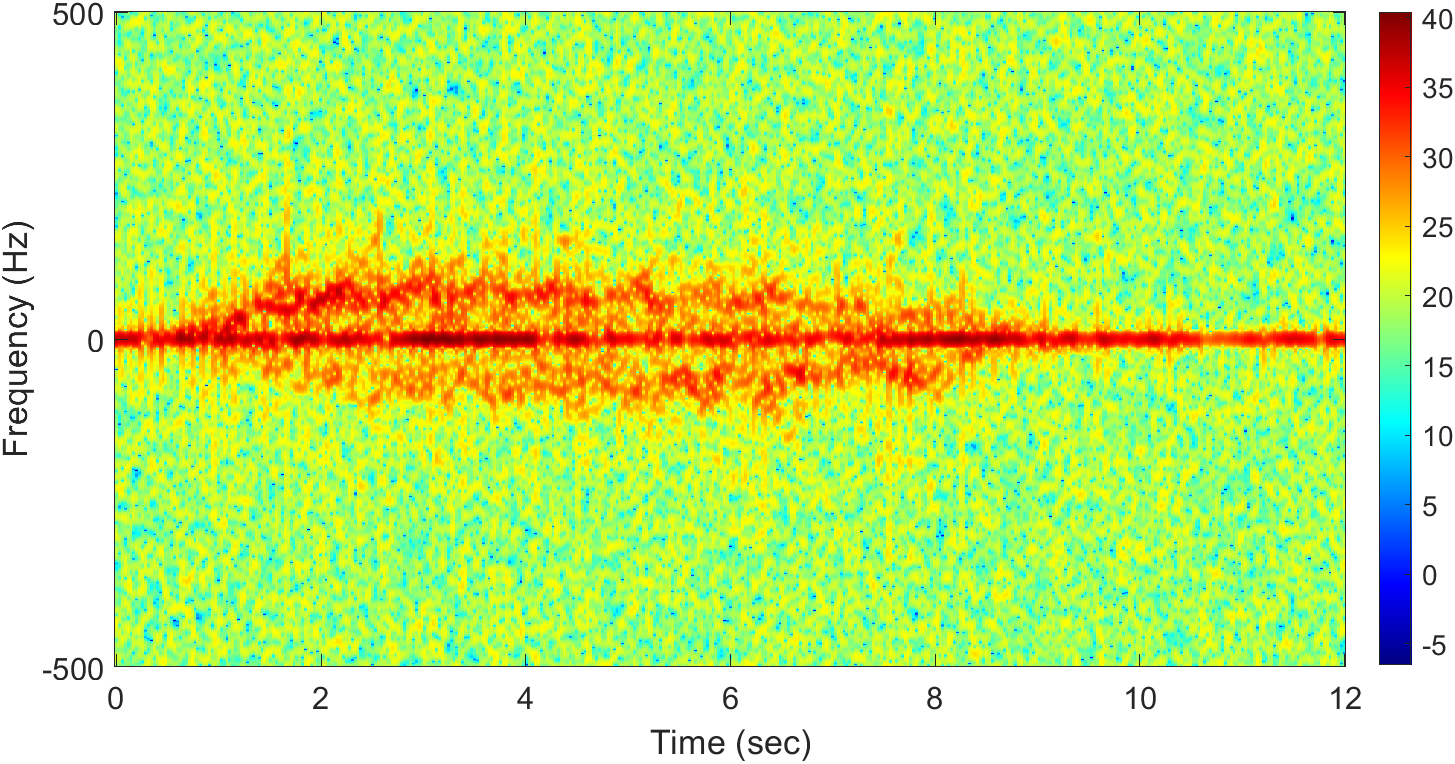}
	\caption{Time Frequency signature of Class-2 motions processed through single antenna (without beamforming)}
	\label{Fig.5}
\end{figure}

\begin{figure}[!t]
	\centering
	\includegraphics[height=2.3in, width=3.5in]{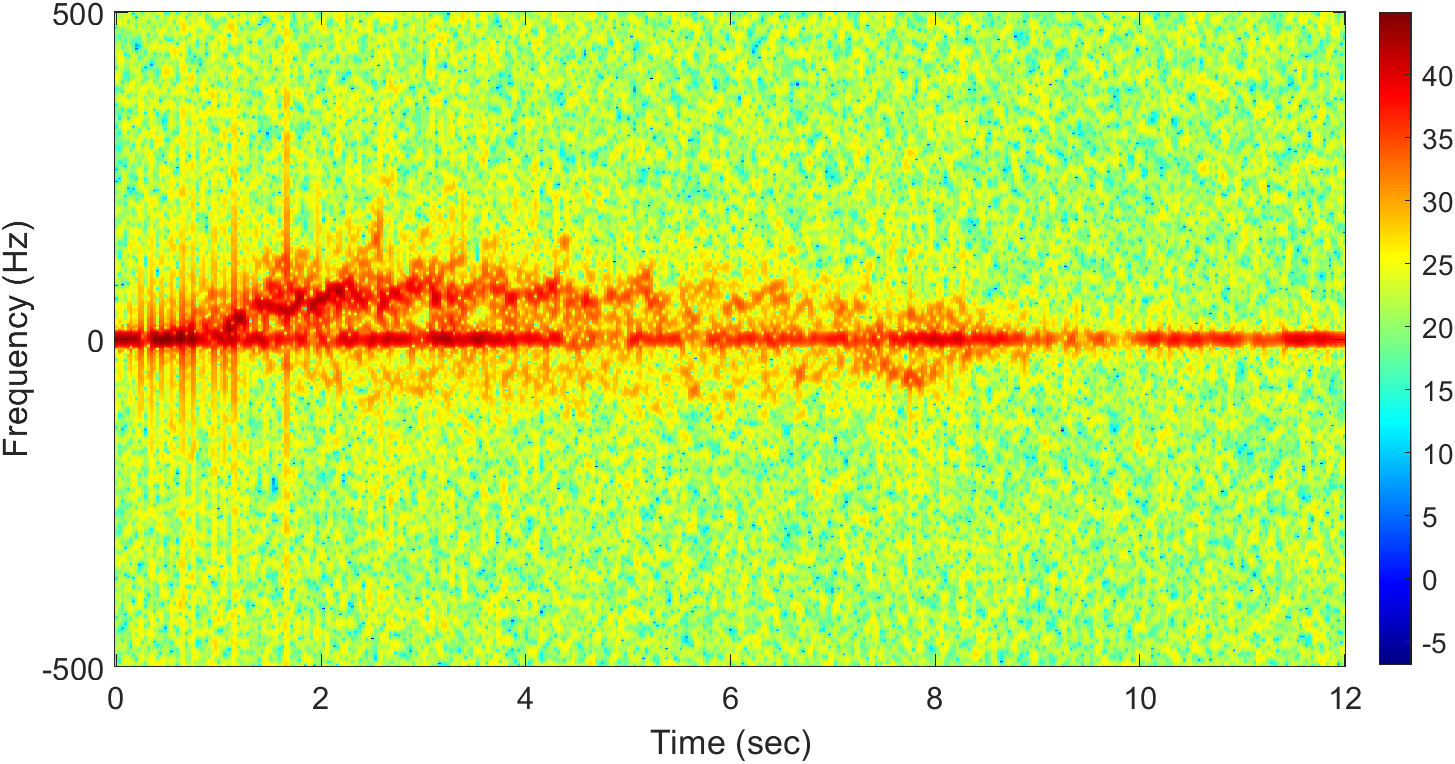}
	\caption{Time Frequency signature of Class-2 motions processed through  beamformer pointing towards $\theta_1$.}
	\label{Fig.6}
\end{figure}

\begin{figure}[!h]
	\centering
	\includegraphics[height=2.3in, width=3.5in]{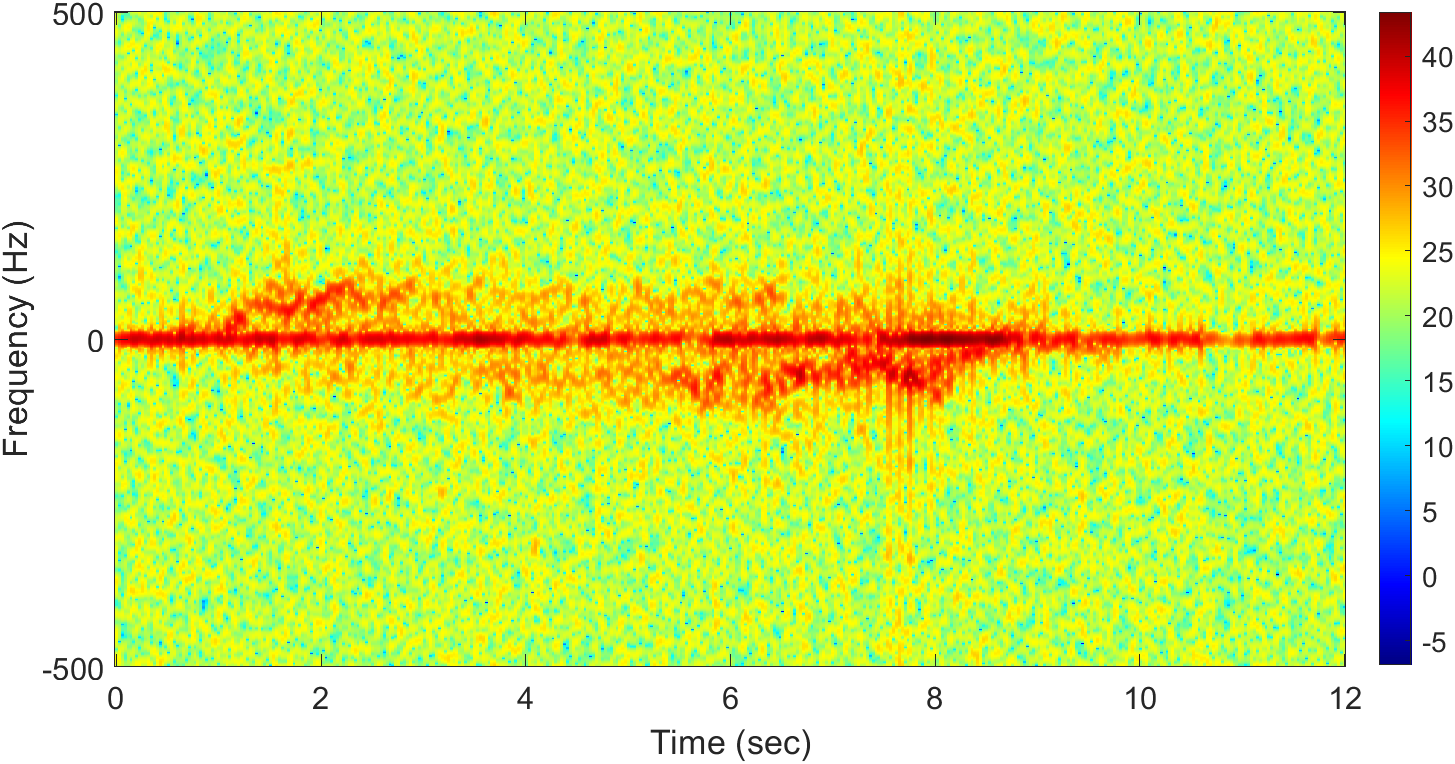}
	\caption{Time Frequency signature of Class-2 motions processed through  beamformer pointing towards $\theta_2$.}
	\label{Fig.7}
\end{figure}

Figure \ref{Fig.2} shows the time frequency signature of  Class-1 motion. Note the TF domain depicts both strong positive and negative Doppler, where both frequencies are present at the same time.   The time frequency signature of  the  Class-1 motion  is processed after  applying beamforming weights to the output of all receivers. The beamformer points to $\theta_1$ direction. Figure \ref{Fig.3} shows the time frequency signature of  the beamformer pointing towards $\theta_1$. It is evident from the  time frequency signature that the beamformer attempted to filter the positive frequencies. However, the positive Doppler is not completely eliminated due to the sidelobes of the beamformer. The beamformer pointing towards $\theta_2$ emphasizes the positive Doppler while attempting to  mitigate the strong negative Doppler   as depicted in Fig. \ref{Fig.4}. On the other hand, Fig. \ref{Fig.5} shows the TF spectrum of the Class-2 motion. It is clear that Figs.  \ref{Fig.2} and \ref{Fig.5} are very similar and are of little use to classify the two motion classes. However, the spatially filtered TF spectrums for Class-2 are flipped for respective look directions when compared to Class-1 motion as shown in Figs. \ref{Fig.6} and \ref{Fig.7}. Therefore, the two beamformed TF spectrums when processed jointly can potentially  classify the two motion classes with reasonable accuracy, therefore mapping the direction of motion to the correct azimuth angle.  


%\subsection{Example 1}
%The FMCW radar experiments were conducted in the EE Lab 333 at Widener University.
The radar system used in the experiments is  AWR2243 from Texas Instruments having four receiver and three transmitters.  The center frequency is 77 GHz, whereas the bandwidth is 5 GHz. The obtained data set is analyzed in MATLAB R2020B. The data set contains 120 samples, 60 samples for each motion class collected through six subjects. For the 2D PCA-based classifications, the training set consists of 46 samples for each class (92 samples in total), and each sample are two 128 × 384 images. The rest of the data set is used for testing. The first image is generated by steering the beamformer to $\theta_1$. Likewise, the other image is generated by changing the steering angle to $\theta_2$. Thus, for each motion class two 128× 384 × 46 matrices  are generated. After performing the dimensionality reduction through 2D PCA,  both  beamformed TF images are concatenated together and passed on to NN classifier. The confusion matrix for fused spectrogram  is shown in Tables \ref{table:1}, depicting the correspondence between the actual and the classified class.    It shows that using only the dominant eigenvector renders a classification accuracy of around $97.6\%$ and  $100\%$ for Class-1 and Class-2 respectively. More specifically, Table \ref{table:1} depicts that the around $97.6\%$ of actual Class-1 motions are classified as Class-1, while the rest $2.4\%$ are declared as Class-2.  The success rate can be improved by including an additional principal component. Table \ref{table:2} shows that the proposed scheme has  a $100\%$ success rate when  two principal eigenvectors are employed.

%\begin{center}

%In this experiment, we consider the case of deceptive jamming. The simulation parameters we used are exactly the same as example A, the only difference is that the angles of interference relative to transmitter and receiver are both $\theta_{q}$. Similarly, we get the relationship curve of output SINR and target angle under different sparsity, as shown in Fig.2. It can be seen from the results that compared with example A, the effect of deceptive interference on the output SINR is much greater when the target is away from the jamming. This is because when the target direction is close to the interference, the matched filter will make the deceptive interference retain more, which will affect the performance of the system.

%\vfill \break

\section{Conclusion}
\label{Conclusion}

In this paper, we introduced an approach that observes the 
time frequency representation of  radar returns from different azimuth angles.  We provided an effective means to discern combinations of multiple motions occurring at different angular directions from the radar. The proposed approach successfully   maps the actual motions to the corresponding angular locations and is  found to be effective when the spectrograms  are not completely separable in angle.  

\section{ACKNOWLEDGMENT}
\label{Problem Formulation}

The authors would like to thank Andrew Lichtenwalner, Daniel Galvao, Ryan Cummings and Connor Ryan for their assistance with data collection.
\bibliographystyle{IEEEtran}
\bibliography{references}
%\end{figure*}

% that's all folks
\end{document}